\documentclass[reprint,aps,prx,superscriptaddress,longbibliography,nofootinbib]{revtex4-2}

\usepackage{xcolor}
\usepackage{graphicx}
\usepackage{amsmath}
\usepackage{amssymb}
\usepackage{hyperref}
\usepackage{physics}

\hypersetup{colorlinks=true, linkcolor=blue, citecolor=blue, urlcolor=blue,
  pdftitle={Generation of entanglement statistics with a large-scale integrated photonic-electronic circuit},
  pdfauthor={Volkan Gurses and Ali Hajimiri}}

\begin{document}

\preprint{}

\title{Generation of entanglement statistics with a large-scale integrated photonic-electronic circuit}

\author{Volkan Gurses}
\thanks{Contact author: gurses@caltech.edu}
\altaffiliation[Present address: ]{Department of Electrical Engineering and Computer Science, Massachusetts Institute of Technology, Cambridge, MA, USA}
\affiliation{Division of Engineering and Applied Science, California Institute of Technology, Pasadena, CA, USA}
\affiliation{Division of Physics, Mathematics and Astronomy, California Institute of Technology, Pasadena, CA, USA}
\affiliation{Institute for Quantum Information and Matter, California Institute of Technology, Pasadena, CA, USA}
\author{Ali Hajimiri}
\affiliation{Division of Engineering and Applied Science, California Institute of Technology, Pasadena, CA, USA}

\date{\today}

\begin{abstract}
Continuous-variable quantum processors scale with the modes they can transform, and each transformation adds loss. We demonstrate optoelectronic quantum information processing, moving the mixing into radio-frequency electronics after homodyne detection. Squeezed vacuum falls on a large-scale integrated photonic-electronic circuit, a 32-channel coherent receiver array whose radio-frequency network mixes the analog photocurrents. A local-oscillator ramp reconstructs the two-mode covariance at the aperture, which is physical, and its partial transpose has smallest symplectic eigenvalue $0.99270$, three standard deviations below the separability bound. Mode transformation becomes a circuit function whose scale follows integrated electronics. With electro-optic transducers in place of the photodiodes, the same receiver would transfer optical modes into superconducting circuits and back, toward a photonic-electronic quantum computer spanning the optical and microwave bands.
\end{abstract}

\maketitle

\section{Introduction}

Encoding quantum information in continuous variables (CV), the quadratures of the electromagnetic field, is one of the two main approaches to optical quantum information processing (QIP) \cite{Weedbrook2012, Lvovsky2009}. Discrete-variable photonic architectures require efficient sources of indistinguishable single photons and scalable entangling operations, whereas CV architectures generate Gaussian resources deterministically and must control errors from finite squeezing and optical loss \cite{Weedbrook2012, masada2015continuous}. Feeding squeezed states from parametric down-conversion \cite{Weedbrook2012} and interfering them in a network of interferometers produces entanglement, usually in the form of cluster states \cite{Weedbrook2012, van2007building, su2007experimental, yukawa2008experimental}, and balanced homodyne detection with a local oscillator (LO) reads out their quadratures \cite{Lvovsky2009}. In measurement-based quantum computation, a sequence of single-mode measurements on a large cluster state with homodyne detection is used \cite{raussendorf2001one, menicucci2006universal}, along with a non-Gaussian measurement, such as photon-number-resolving detection, to enable universal quantum computation \cite{menicucci2006universal}. The same squeezed states and linear optical networks underlie Gaussian boson sampling, which samples the photon-number statistics of squeezed states sent through a large interferometer and has been implemented on a programmable photonic processor \cite{arrazola2021quantum} and at the scale of quantum computational advantage \cite{madsen2022quantum}. The same scheme also enables CV quantum communications, which transmits coherent or squeezed states to a homodyne receiver \cite{grosshans2002continuous, Weedbrook2012, GursesOFC2024, Gurses2025, GursesAPS2025, DavisAPS2025, GursesQCT2026}, and quantum sensing, which injects squeezed vacuum into the interferometer to improve the sensitivity of shot-noise-limited sensors \cite{tse2019quantum, Gurses2025, GursesAPS2025, DavisAPS2025}.

The number of entangled modes and their entanglement structure determine the resource available for computation \cite{Yokoyama2013}. Scaling CV processors has so far required scaling the number of modes in linear optical networks. In a spatially multiplexed network, a fixed layout of beam splitters and phase shifters transforms the input modes into entangled modes, a different set of entangled modes requires a different network \cite{armstrong2012programmable, ferrini2013compact}, and every additional mode adds components and complexity to the network \cite{Yokoyama2013}. Time- and frequency-domain multiplexing reuse the same optical components for each additional mode, separated in time or in frequency \cite{Yokoyama2013, Roslund2014, Asavanant2019}, which makes the component count independent of the circuit size \cite{madsen2022quantum}, and the delay lines and mode-resolving local oscillators that they require add loss and phase drift before detection. Integrated photonics provides the phase stability and mode matching that spatially multiplexed networks of many interferometers require \cite{Harris2017, arrazola2021quantum}, and integrated CV photonics has advanced from two-mode entanglement on a chip \cite{masada2015continuous} to eight-mode entanglement in a microcomb \cite{Jia2025}, to monolithic generation, manipulation, and measurement of cluster states \cite{Jia2026}, and to a modular architecture for fault-tolerant photonic quantum computing \cite{AghaeeRad2025}. In free space, spatial light modulators, metasurfaces, and optical phased arrays transform the spatial modes of the field in the optical domain \cite{panuski2022full, Yousef2025, GursesNatPhot2022}. In each of these approaches the mode transformation is applied to the quantum field before detection, so the propagation loss, crosstalk, and thermal drift of the network degrade the state \cite{gurses2022large, Clark2026}, and optical loss remains the main obstacle to fault tolerance \cite{AghaeeRad2025}.

Electronic noise in homodyne detection is equivalent to optical loss when the detector is calibrated against vacuum \cite{appel2007electronic}, so the shot noise of the signal field must be resolved well above the electronic noise of the detector, and their ratio, the shot-noise clearance, sets the equivalent efficiency \cite{GursesFiO2022, GursesOFC2024, GursesQCT2026, Bruynsteen2021, Tasker2021}. The detector bandwidth defines the clock rate of a CV quantum computer \cite{Tasker2021}. The first homodyne detector integrated in silicon photonics reached 150~MHz \cite{raffaelli2018homodyne}, since the transimpedance amplifier (TIA) that amplifies the weak photocurrent difference was a discrete circuit whose parasitic capacitance adds noise and limits bandwidth \cite{Tasker2021}, and interfacing balanced germanium photodiodes directly with an integrated TIA has extended shot-noise-limited operation to tens of gigahertz \cite{GursesOFC2024, Tasker2021, Bruynsteen2021}. Such a co-integrated receiver is a quantum-limited coherent receiver, and we have demonstrated a 32-channel array of them with $30.3$~dB of shot-noise clearance \cite{Gurses2023, GursesAPS2023, Gurses2025, GursesCLEO2025}.

In this work, we introduce optoelectronic quantum information processing (Fig.~\ref{fig:concept}), which moves certain transformations of a circuit from photonics to electronics. We demonstrate it with squeezed vacuum on a large-scale integrated photonic-electronic circuit, a 32-channel silicon-photonic quantum-limited coherent receiver array with on-chip thermo-optic phase shifters and balanced germanium photodiodes feeding RF hybrids and digital processing, with eight channels active in the reported acquisition. The RF hybrids carry out the transformation that takes the detected channels to the two modes of a two-mode cluster state, an LO ramp reconstructs its two-mode covariance, and digital processing returns its entanglement statistics. The reconstructed covariance matrix corresponds to a physical state and violates the positive-partial-transpose (PPT) criterion. The certification concerns the optical modes selected from the field delivered to the aperture. The electronic processing leaves no optical output state. Electronic linear processing is reprogrammable and free of interferometric drift, and electronic loss after detection does not attenuate the optical state. Off-loading a photonic circuit function in this way removes the loss that the same transformation adds in the optical domain, and the mixing layer then scales with integrated electronics rather than with the depth of the optical network. A cryogenic quantum-limited coherent receiver co-located with superconducting circuits would deliver quantum-noise-resolved quadrature statistics directly to superconducting qubits, and with electro-optic transducers in place of its photodiodes and the LO as their pump it would transfer the optical modes themselves into microwave modes. Optoelectronic quantum circuits of this kind, in which integrated photonics generates and distributes squeezing and entanglement and integrated electronics carries out the mode transformations, the measurement processing, and the feedforward of the protocol, could be a route toward photonic-electronic quantum computers, transducers, and networks.

\section{Electronic mode transformation}
\label{sec:framework}

\begin{figure*}[t]
  \centering
  \includegraphics[width=\textwidth]{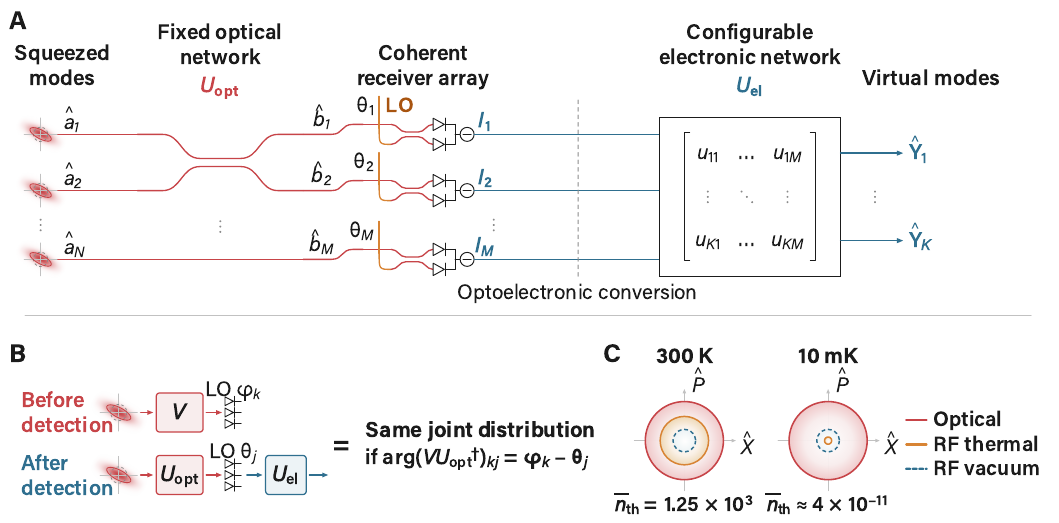}
  \caption{\textbf{Optoelectronic quantum information processing.} \textbf{(A)} Electronic mode transformation. A multimode squeezed field propagates through the fixed optical network $U_\text{opt}$ to a coherent receiver array whose LO phases realize $D(\vec{\theta})$, and a configurable real electronic network $U_\text{el}$ combines the photocurrents into output quadratures $\hat{Y}_k$. With $W=U_\text{el}D(\vec{\theta})U_\text{opt}$, each $\hat{Y}_k$ has the homodyne statistics of the normalized virtual mode $\hat{c}_k$, scaled by $\norm{\vec{w}_k}$ [Eq.~(\ref{eq:measurement})]. \textbf{(B)} Measurement equivalence. A target optical network $V$ before detection, read out at LO phases $\varphi_k$, and the installed network $U_\text{opt}$ followed by detection at LO phases $\theta_j$ and a configurable real electronic network $U_\text{el}$ return the same joint distribution when the target rows lie in the detected subspace and Eq.~(\ref{eq:condition}) holds. \textbf{(C)} Optical readout and microwave noise scales (schematic, radii not to scale). The $30.3$-dB single-channel clearance is measured in the receiver characterization band and places the LO-amplified shot noise of the signal field above electronic noise. The separate microwave comparison is for a $5$-GHz mode, whose thermal occupation is large at room temperature and negligible near $10$~mK.}
  \label{fig:concept}
\end{figure*}

We consider a multimode input field with annihilation operators $\vec{a}_\text{in} = (\hat{a}_1, \hat{a}_2, \ldots, \hat{a}_N)^T$, $[\hat{a}_n, \hat{a}_m^\dagger] = \delta_{nm}$, in any orthonormal basis of spatial, temporal, or frequency modes. A fixed linear optical transformation $U_\text{opt} \in \mathbb{C}^{M \times N}$ with orthonormal rows, $U_\text{opt}U_\text{opt}^\dagger = \openone_M$, routes the field to an array of $M$ quantum-limited coherent receivers with LO phases $\{\theta_j\}$. Receiver $j$ measures the quadrature
\begin{equation}
    \hat{X}_j(\theta_j) = \frac{1}{\sqrt{2}}\left(\hat{b}_j e^{-i\theta_j} + \hat{b}_j^\dagger e^{i\theta_j}\right)
    \label{eq:quadrature}
\end{equation}
of the mode $\hat{b}_j = \sum_n (U_\text{opt})_{jn}\, \hat{a}_n$ coupled into it and produces a photocurrent $I_j(t) \propto \hat{X}_j(\theta_j)$, LO-amplified above the electronic noise of the readout chain \cite{appel2007electronic, Gurses2025}. Throughout, $\hat{X}_c = (\hat{c} + \hat{c}^\dagger)/\sqrt{2}$ and $\hat{P}_c = (\hat{c} - \hat{c}^\dagger)/(i\sqrt{2})$, so that $[\hat{X}_c, \hat{P}_c] = i$ and each quadrature has vacuum variance $1/2$. The modes $\hat{b}_j$ are mutually orthogonal, so the operators $\{\hat{X}_j(\theta_j)\}$ commute and are measured jointly in a single shot.

An electronic circuit, analog or digital, acting on the photocurrent vector $\vec{I}(t)$ with a real matrix $U_\text{el} \in \mathbb{R}^{K \times M}$ synthesizes the observables
\begin{equation}
    \hat{Y}_k = \sum_{j=1}^{M} (U_\text{el})_{kj}\, \hat{X}_j(\theta_j).
    \label{eq:electronic}
\end{equation}
With $D(\vec{\theta}) = \text{diag}(e^{-i\theta_1}, \ldots, e^{-i\theta_M})$,
\begin{equation}
    W = U_\text{el}\, D(\vec{\theta})\, U_\text{opt},
    \label{eq:W}
\end{equation}
and $\vec{w}_k$ the $k$th row of $W$, a direct calculation (Appendix~\ref{app:derivation}) gives
\begin{equation}
    \hat{Y}_k = \norm{\vec{w}_k}\, \hat{X}_{c_k},
    \qquad
    \hat{c}_k = \frac{1}{\norm{\vec{w}_k}} \sum_n W_{kn}\, \hat{a}_n,
    \label{eq:measurement}
\end{equation}
where $\norm{\vec{w}_k} = \big(\sum_n |W_{kn}|^2\big)^{1/2}$, so that $[\hat{c}_k, \hat{c}_k^\dagger] = 1$. A real $U_\text{el}$ makes the coefficient of $\hat{a}_n^\dagger$ in Eq.~(\ref{eq:expand}) the complex conjugate of the coefficient of $\hat{a}_n$, whereas for complex $U_\text{el}$ the combination is not Hermitian and is not an observable.

Each output quadrature $\hat{Y}_k$ has the statistics of a homodyne measurement of the virtual mode $\hat{c}_k$, up to the scale $\norm{\vec{w}_k}$, which vacuum calibration removes. The virtual modes are orthonormal when the rows of $W$ are orthogonal, that is, when $WW^\dagger$ is diagonal, as in the experiment below. The $\hat{Y}_k$ commute whether or not the virtual modes are orthogonal, so one acquisition gives their joint distribution, whereas conjugate quadratures of one virtual mode require separate settings of $\vec{\theta}$ or $U_\text{el}$. For a densely sampled aperture the rows of $W$ approach continuous mode functions, and Eq.~(\ref{eq:measurement}) becomes electronically programmed mode-selective homodyne detection \cite{fabre2020modes, Roslund2014}.

Equation~(\ref{eq:measurement}) determines the accessible Gaussian circuits. Let $V\in\mathbb{C}^{K\times N}$ contain the $K$ measured rows of a target passive transformation, an optical network distinct from the network $U_\text{opt}$ installed in the path, with $VV^\dagger=\openone_K$. For arbitrary input states, the receiver reproduces the statistics of applying $V$ and measuring its outputs at homodyne angles $\varphi_k$ if and only if (i) the measured target rows have no support outside the detected subspace, $D(\vec{\varphi})\, V \left(\openone_N - U_\text{opt}^\dagger U_\text{opt}\right) = 0$, and (ii) $D(\vec{\varphi})\, V\, U_\text{opt}^\dagger\, D(\vec{\theta})^\dagger$ is real,
\begin{equation}
    \arg\!\left[\left(V U_\text{opt}^\dagger\right)_{kj}\right] = \varphi_k - \theta_j \pmod{\pi}
    \label{eq:condition}
\end{equation}
for every nonzero entry $\left(V U_\text{opt}^\dagger\right)_{kj}$. The electronic map is then $U_\text{el} = \Lambda\, D(\vec{\varphi})\, V\, U_\text{opt}^\dagger\, D(\vec{\theta})^\dagger$ for any diagonal matrix $\Lambda$ of positive row gains, or equivalently $U_\text{el}\, D(\vec{\theta})\, U_\text{opt} = \Lambda\, D(\vec{\varphi})\, V$, so the electronic network implements the part of the target transformation that the installed optics does not [Fig.~\ref{fig:concept}(B)]. A complex gain would change the target homodyne angle. For a full $N$-output unitary, condition (i) requires detection of all $N$ input modes, and the row form also covers protocols that measure a supported subset. We call a target measurement compatible when both the support requirement and Eq.~(\ref{eq:condition}) hold. Equation~(\ref{eq:condition}) expresses the multipixel feasibility condition of Ref.~\cite{ferrini2013compact} for specified terminal homodyne measurements and makes the support requirement in a partially detected mode space explicit. In the fully detected square case it reduces to that condition after the target homodyne phases are included in the target matrix. Reconfigurable optics \cite{gurses2022large, GursesNatPhot2022}, sequential measurements, and feedforward \cite{larsen2021deterministic} expand the accessible set. Squeezing enters through the optical input and displacements through feedforward. Universal quantum computation requires in addition a nonlinear element, which a non-Gaussian measurement supplies \cite{menicucci2006universal, Weedbrook2012}.

Moving the transformation after detection costs receiver noise, which is quantified by the shot-noise clearance $\mathcal{C}$, the ratio of the shot noise of the signal field to the electronic noise at the measurement sideband frequency \cite{GursesFiO2022, GursesOFC2024, Gurses2025, GursesQCT2026, Bruynsteen2021, Tasker2021}. Input-referred electronic noise is equivalent to an optical loss $1 - \eta_\text{el}$ with $\eta_\text{el} = (1 + \mathcal{C}^{-1})^{-1}$ \cite{appel2007electronic}, which gives $\eta_\text{el} = 0.9991$ at the measured $30.3$-dB single-channel clearance. Noise added after the high-gain front end is reduced by the preceding gain when referred to the input, so electronic network depth adds no optical loss, whereas every layer of a linear optical network attenuates the state \cite{Harris2017, Clark2026}.

Equation~(\ref{eq:measurement}) establishes an equivalence between measurement statistics and does not describe a state of the electronic outputs. After detection, the electronics selects the virtual mode basis and the terminal homodyne setting that the output quadrature represents. Complementary output quadratures combine into the same Duan or cluster-nullifier function as complementary pre-homodyne settings, and a violation of its bound is an entanglement statistic of the emulated circuit and is not evidence that photodetection created or preserved an entangled electronic output.

\section{Experimental implementation}
\label{sec:implementation}

\begin{figure*}[t]
  \centering
  \includegraphics[width=\textwidth]{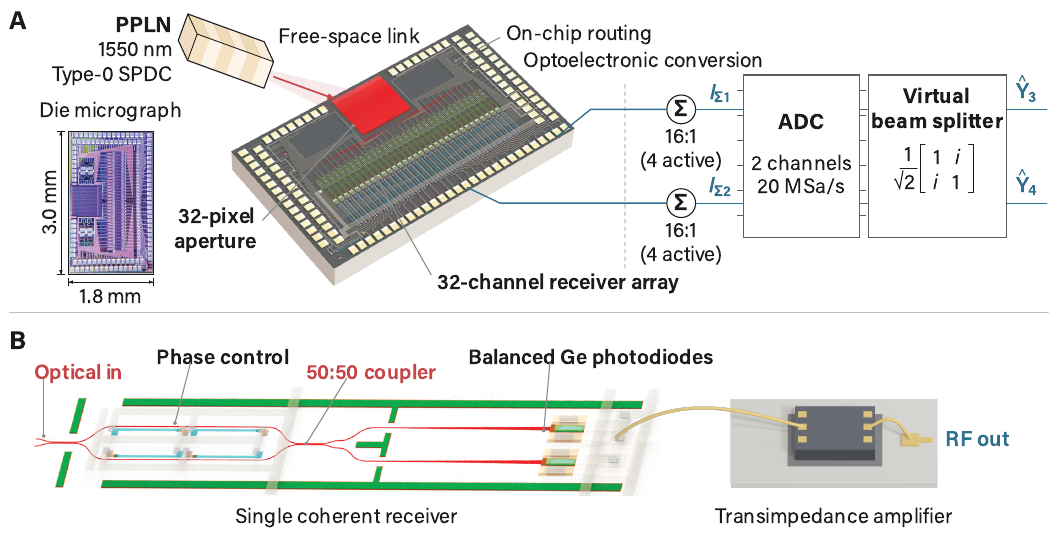}
  \caption{\textbf{Experimental implementation.} \textbf{(A)} The measurement chain. Broadband squeezed vacuum at $1550$~nm from type-0 spontaneous parametric down-conversion (SPDC) in periodically poled lithium niobate (PPLN) crosses a free-space link onto the 32-pixel metamaterial aperture (red) of the silicon-photonic receiver die, each antenna element feeding one channel of the 32-channel coherent receiver array. The four channels nearest the array center are retained on each 16-channel half, so each 16:1 RF hybrid network carries four active inputs. The two sums are digitized by a two-channel analog-to-digital converter (ADC) at $20$~MSa/s. \textbf{(B)} Signal chain of one receiver, with optical input, thermo-optic phase control $\theta_j$, 50:50 coupler, balanced germanium photodiodes, and the transimpedance amplifier (TIA) driving the hybrid network. The inset is a micrograph of the $3.0$ by $1.8$~mm die. The die rendering, die micrograph, and layout artwork in (A) and (B) are adapted from Ref.~\cite{Gurses2025}, \textcopyright{} The Author(s) 2025, published under a Creative Commons Attribution 4.0 International License.}
  \label{fig:impl}
\end{figure*}

We implement a two-mode CV cluster-state protocol \cite{armstrong2012programmable, ferrini2013compact, van2007building, yukawa2008experimental} with a 32-channel silicon-photonic quantum-limited coherent receiver array followed by RF hybrids and digital processing (Fig.~\ref{fig:impl}) \cite{Gurses2025, GursesCLEO2025}. Type-0 spontaneous parametric down-conversion in periodically poled lithium niobate generates broadband squeezed vacuum at $1550$~nm, which illuminates the receiver chip over a free-space link. Each receiver contains a tunable Mach--Zehnder interferometer, balanced germanium photodiodes, and a TIA. On-chip thermo-optic phase shifters set the per-channel LO phase $\theta_j$ \cite{gurses2022large}, and a common linear phase ramp scans all channels. Single-channel characterization gives $30.3$~dB shot-noise clearance and a time-averaged common-mode rejection ratio of $90.2$~dB at $1.1$~MHz \cite{Gurses2025}.

The photocurrents of each half of the array are summed by a 16:1 RF hybrid network. For the inseparability acquisition, the four channels nearest the array center are retained on each half and the other twelve are pruned, since an unilluminated channel contributes vacuum noise but negligible signal to the sum, and the eight-channel configuration improves the geometric overlap efficiency of the combined mode. The RF hybrid network implements $G \oplus G$, where $G$ is the row vector with unit entries on the four retained channels of one half and zeros on the other twelve. The two combined photocurrents are digitized at $20$~megasamples per second (MSa/s). We label the virtual modes of the two combined output quadratures 1 and 2. The target circuit is the beam splitter
\begin{equation}
    S = \frac{1}{\sqrt{2}}
    \begin{pmatrix}
    1 & i \\
    i & 1
    \end{pmatrix}
    \label{eq:S}
\end{equation}
with the complex entries realized as $\pi/2$ quadrature rotations at the analysis frequency and with its outputs labeled 3 and 4. The squeezed vacuum is a single mode, so the state shared by the two halves is the output of a beam splitter fed by one squeezed mode and vacuum \cite{kim2002entanglement}, and the two-mode cluster state is the target of the protocol whose nullifiers we test. The full chain
\begin{equation}
    W_\text{exp} = S\,(G \oplus G)\, D(\vec{\theta})\, U_\text{opt}
    \label{eq:Wexp}
\end{equation}
has the form of Eq.~(\ref{eq:W}), with the hardware-fixed basis change $U_\text{opt}$ (free-space propagation and on-chip routing) and programmable LO phases $D(\vec{\theta})$. The reported acquisition realizes $(G \oplus G)\, D(\vec{\theta})\, U_\text{opt}$. The two-mode covariance reconstructed in Sec.~\ref{sec:results} is that of the two combined output quadratures before $S$ (Appendix~\ref{app:stats}), and the covariance of the outputs of $S$ follows from it by a change of basis.

The electronic map $A = S\,(G \oplus G)$ is complex, whereas Eq.~(\ref{eq:electronic}) requires a real matrix. At analysis frequency $\Omega$, let $\hat{a}_{j\pm}$ denote the sidebands at $\omega_0\pm\Omega$, with the symmetric and antisymmetric sideband modes $\hat{s}_j=(\hat{a}_{j+}+\hat{a}_{j-})/\sqrt{2}$ and $\hat{d}_j=(\hat{a}_{j+}-\hat{a}_{j-})/\sqrt{2}$ \cite{caves1985two}. Up to convention-dependent signs and a common scale, the in-phase (I) and quadrature (Q) components of the photocurrent measure $\hat{X}_{s_j}(\theta_j)$ and $\hat{P}_{d_j}(\theta_j)$, and on the stacked I and Q components the complex filter $A$ has the real representation
\begin{equation}
    \mathcal{R}(A) =
    \begin{pmatrix}
        \operatorname{Re}A & -\operatorname{Im}A \\
        \operatorname{Im}A & \phantom{-}\operatorname{Re}A
    \end{pmatrix}.
    \label{eq:realification}
\end{equation}
Because $S$ is unitary and the two rows of $G\oplus G$ have disjoint support and equal norm, $\mathcal{R}(A)\mathcal{R}(A)^T=\norm{G}^{2}\openone_4$. Equation~(\ref{eq:measurement}) therefore applies to the four real outputs on the space of the symmetric and antisymmetric sideband modes of the two halves, on which the fixed optical map acts as $U_\text{opt}\oplus U_\text{opt}$, again with orthonormal rows, at measured angles $\theta_j$ and $\theta_j+\pi/2$. The I and Q output quadratures emulate complementary pre-homodyne settings without retuning the LO.

The target $V = S$ also satisfies Eq.~(\ref{eq:condition}) on the symmetric sideband modes alone, without any digital rotation. With $\theta_j = \theta$ on one half of the array and $\theta_j = \theta + \pi/2$ on the other, the real matrix
\begin{equation}
    U_\text{el} = \frac{1}{\sqrt{2}}
    \begin{pmatrix}
    G & -G \\
    G & G
    \end{pmatrix} \in \mathbb{R}^{2 \times 32}
    \label{eq:Mreal}
\end{equation}
synthesizes the commuting pair $\hat{X}_3$ and $\hat{P}_4$ from the thermo-optic phase shifters and the sum and difference of the two combined photocurrents, and a complementary setting resolves the other two output quadratures, so that the two-mode functions of Sec.~\ref{sec:results} can be assembled across acquisitions. Electronic phase tuning of the I/Q output quadratures emulates both settings from one acquisition. The two routes return the same statistic when the symmetric and antisymmetric sideband modes carry equal covariances, $\text{Var}\,\hat{X}_{s_j} = \text{Var}\,\hat{P}_{d_j}$, with no correlation between them, which holds for the degenerate parametric source used here, and when the two array halves couple with equal overlap efficiency (Appendix~\ref{app:stats}).

\section{Certified inseparability of the two-mode cluster state}
\label{sec:results}

The nullifiers of an ideal cluster state are linear combinations of position and momentum operators with eigenvalue zero \cite{Gu2009}. Finite squeezing gives nonzero nullifier variances \cite{van2007building, Gu2009}. For the two-mode cluster state considered here the nullifiers are $\hat{X}_1 - \hat{P}_2$ and $\hat{X}_2 - \hat{P}_1$, the $\hat{p}_1 - \hat{x}_2$ and $\hat{p}_2 - \hat{x}_1$ of Refs.~\cite{van2007building, Gu2009} after a local phase rotation of each mode, which the beam splitter maps onto the $X$ quadratures of its outputs,
\begin{equation}
    \sqrt{2}\,\hat{X}_3 = \hat{X}_1 - \hat{P}_2,
    \qquad
    \sqrt{2}\,\hat{X}_4 = \hat{X}_2 - \hat{P}_1.
    \label{eq:nullifiers}
\end{equation}
We evaluate two entanglement criteria from the same reconstructed covariance matrix, the PPT criterion and the variances of the locally phase-optimized nullifiers normalized to vacuum \cite{su2007experimental, yukawa2008experimental}.

Within the Gaussian source model, the second moments collected in the covariance matrix define the state up to displacements \cite{Weedbrook2012, adesso2007entanglement}, so we characterize the two virtual modes by measuring the entries of their two-mode covariance matrix $\Sigma$. For $\vec{R}=(\hat{X}_1,\hat{P}_1,\hat{X}_2,\hat{P}_2)^T$ and $\Delta\hat{R}_\mu=\hat{R}_\mu-\langle\hat{R}_\mu\rangle$, we define $\Sigma_{\mu\nu}=\langle\Delta\hat{R}_\mu\Delta\hat{R}_\nu+\Delta\hat{R}_\nu\Delta\hat{R}_\mu\rangle$. This gives $\Sigma_\text{vac}=\openone_4$ with the quadrature convention of Eq.~(\ref{eq:quadrature}). The vacuum noise reference is measured with the squeezed beam blocked (Appendix~\ref{app:stats}). The LO scan rotates the measurement basis of each sideband mode as $\hat{X}_{s}(\theta) = \hat{X}_{s}\cos\theta + \hat{P}_{s}\sin\theta$ [Eq.~(\ref{eq:quadrature})], so that every second moment of the output quadratures is a sinusoid in $\psi \equiv 2\theta$,
\begin{equation}
    m(\psi) = a + b\cos\psi + c\sin\psi,
    \label{eq:tomo}
\end{equation}
whose fitted coefficients determine the quadrature variances and covariances. The ramp constrains nine independent linear combinations of the ten entries of $\Sigma$, and the remaining one, the antisymmetric part of the cross-correlation between $\hat{X}$ and $\hat{P}$ of the two modes, is set to zero, which changes $\tilde\nu_-$ by less than $10^{-4}$ over the range of that entry for which $\Sigma$ remains a physical covariance matrix (Appendix~\ref{app:stats}). The antisymmetric sideband modes yield a second, physically distinct two-mode covariance matrix by the same fit.

\begin{figure*}[t]
  \centering
  \includegraphics[width=\textwidth]{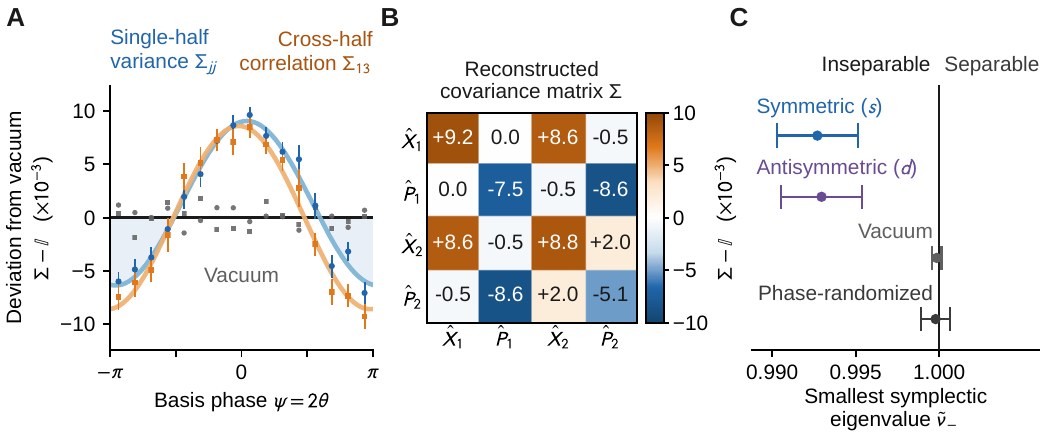}
  \caption{\textbf{Inseparability of the two-mode cluster state.} \textbf{(A)} Second moments of the sideband-mode quadratures during the LO ramp, binned in the recovered phase $\psi = 2\theta$ and plotted as deviations from vacuum in units of $10^{-3}$, for the mean single-half variance $\Sigma_{jj}=(\Sigma_{11}+\Sigma_{33})/2$ (blue circles) and the cross-half correlation $\Sigma_{13}$ after the electronic-offset correction (orange squares). Lines are the fits of Eq.~(\ref{eq:tomo}) and error bars are $\pm1$ standard error per bin. The shaded wedge marks where the fitted variance falls below vacuum, and gray points are the vacuum acquisition. \textbf{(B)} Reconstructed two-mode covariance matrix of the two-mode cluster state for the symmetric sideband modes, ordered $(\hat{X}_1,\hat{P}_1,\hat{X}_2,\hat{P}_2)$, with the vacuum contribution subtracted as in (A), blue below vacuum and orange above. The cross-quadrature entries $\Sigma_{14}$ and $\Sigma_{23}$ are constrained to be equal because the ramp resolves their sum but not their difference. \textbf{(C)} Smallest partial-transpose symplectic eigenvalue $\tilde\nu_-$ for the symmetric ($s$) and antisymmetric ($d$) sideband modes, for the vacuum acquisition through the same analysis, and for the mean of the phase-randomized surrogate null. Separability requires $\tilde\nu_-\geq1$ (black line). Error bars for the two sideband-mode pairs combine the moving-block bootstrap standard deviation ($0.0008$) and the resolution-based vacuum-calibration allowance ($0.0023$) in quadrature. The vacuum bar is $\pm1$ bootstrap s.d. and the null bar is $\pm1$ s.d. of the $400$ surrogates (Appendix~\ref{app:stats}).}
  \label{fig:data}
\end{figure*}

The phase $\psi$ is recovered from the power of the two combined photocurrents in reference bands disjoint from the $3$--$6$~MHz test bands that enter $\Sigma$. The measured cross-half statistics contain an electronic offset, a component that does not rotate with the recovered optical basis and is subtracted before the covariance matrix is assembled (Appendix~\ref{app:stats}). The remaining cross-half correlation has amplitude $0.00860$ and follows the single-half variance, whose fitted amplitude is $0.00772$, with binned correlation $+0.973$ and a ratio of cross-half to single-half modulation amplitude of $1.11$ [Fig.~\ref{fig:data}(A)]. With ordinary vacuum replacing the squeezed resource, the binned correlation is $+0.044$ and the amplitude ratio $0.62$, consistent with no shared modulation.

The state condition $\Sigma + i\Omega \geq 0$, with $\Omega$ the two-mode symplectic form, determines whether the reconstructed covariance matrix corresponds to a physical state, and in these units it reads $\nu_-\geq1$ for the smallest symplectic eigenvalue of $\Sigma$ \cite{simon2000peres, adesso2007entanglement}. The reconstructed covariance matrix has $\nu_-=1.00078$ for the symmetric and $1.00092$ for the antisymmetric sideband modes and is therefore a physical covariance matrix, whereas before the electronic-offset correction the uncorrected second-moment matrix has $\nu_-=0.8543$ and does not correspond to a physical state. Partial transposition of one mode of a separable state yields a physical state with symplectic eigenvalues greater than or equal to one, so a smallest symplectic eigenvalue $\tilde\nu_- < 1$ of the partially transposed covariance matrix proves the state inseparable, for Gaussian and non-Gaussian states alike \cite{simon2000peres, werner2001bound}. In the local readout model, loss and uncorrelated readout noise act separately on each half and cannot create entanglement, so $\tilde\nu_- < 1$ for the detected covariance matrix implies inseparability of the field upstream of detection. Tracing out the antisymmetric sideband modes of each half is likewise a local operation, so a violation for the symmetric sideband modes alone suffices.

For the reported band assignment, the symmetric and antisymmetric sideband modes give
\begin{equation}
\begin{aligned}
    \tilde\nu_-^{(s)} &= 0.99270\;[0.99130,\,0.99450], \\
    \tilde\nu_-^{(d)} &= 0.99295\;[0.99162,\,0.99446],
\end{aligned}
    \label{eq:result}
\end{equation}
with $95\%$ moving-block bootstrap percentile intervals in brackets, and a $\pm0.01$~dB systematic allowance on the LO-power correction between the squeezed and vacuum acquisitions gives the calibration uncertainty (Appendix~\ref{app:stats}). The result is $\tilde\nu_-^{(s)} = 0.99270 \pm 0.00081$~(stat) $\pm\,0.00229$~(cal), giving a nominal $3.0\sigma$ margin below unity, or $0.032$~dB below the separability bound. The amount of entanglement is quantified by the logarithmic negativity, $E_N = \max\{0, -\ln\tilde\nu_-\} = 0.0073$ within a Gaussian model \cite{vidal2002computable, adesso2007entanglement}. The two results correspond to physically distinct modes from one acquisition, share the vacuum calibration and recovered phase reference, and agree to $2.5\times10^{-4}$. Applying $S$ to the reconstructed covariance matrix gives $\tilde\nu_- = 0.9936$ for the partition of its two outputs, with $\nu_- = 1.00078$ unchanged.

The vacuum-normalized nullifier variances are evaluated at one common pair of local phases chosen to minimize the larger of the two variances. Both nullifier variances are $0.99275$ for the symmetric and $0.99300$ for the antisymmetric sideband modes, below the vacuum level, against $0.99997$ with vacuum at the input. Their sums, $1.9855$ and $1.9860$ against the separability bound of $2$, violate the sum inequality of Duan et al. \cite{duan2000inseparability}, the two-mode case of the van Loock--Furusawa criterion \cite{vanloock2003detecting}. These values track $\tilde\nu_-$ to $5\times10^{-5}$ and carry the same statistical and calibration uncertainties.

With vacuum replacing the squeezed resource, the identical analysis returns $\tilde\nu_- = 0.99985$, a departure from unity about fifty times smaller than the observed violation. The phase-randomized surrogate distribution has mean $0.99978$, with the observation $8.3$ surrogate standard deviations below that mean. This standardized distance is a diagnostic of phase alignment, not a calibrated significance against all separable states. All $151$ assignments of the band-scan family yield physical covariance matrices and lie below the separability bound, spanning $0.99204$ to $0.99456$ with median $0.99336$ (Appendix~\ref{app:stats}).

In a Gaussian loss model in which every imperfection is equivalent to an additional loss $1-\eta$ at the source of the squeezed light \cite{Asavanant2019}, the two halves invert to squeezing parameters $r = 0.10$ and $0.26$ with total efficiencies $\eta = 4.1\%$ and $1.3\%$, limited by free-space coupling, beam overlap with the retained channels, on-chip routing loss, and photodiode quantum efficiency. The inversion is ill-conditioned and these are model-dependent estimates (Appendix~\ref{app:stats}). The equivalent electronic loss $1-\eta_\text{el} = 9.32\times10^{-4}$ is more than an order of magnitude below the inferred optical loss.

\section{Discussion and outlook}

We have demonstrated optoelectronic QIP, in which a passive Gaussian transformation is moved from a linear optical network into RF electronics after coherent detection, by generating the entanglement statistics of a two-mode cluster state from squeezed vacuum on eight channels of a 32-channel silicon-photonic quantum-limited coherent receiver array and certifying its inseparability by the PPT criterion, $\tilde\nu_- = 0.99270 \pm 0.00081$~(stat) $\pm\,0.00229$~(cal), a nominal $3.0\sigma$ margin below the separability bound. Equation~(\ref{eq:condition}) states when the substitution is exact for a receiver whose optics are fixed and whose homodyne angles are set by the protocol, and the substitution costs receiver noise equivalent to an optical loss $1 - \eta_\text{el}=(1+\mathcal{C})^{-1}$, which is incurred once at detection rather than once per layer of the circuit.

Electronic sums of homodyne photocurrents with adjustable gains have measured cluster-state nullifiers since the first CV cluster-state experiments \cite{su2007experimental, yukawa2008experimental}. The closest previous demonstration programmed virtual linear optical networks by weighting the per-pixel photocurrents of an eight-pixel homodyne detector numerically on a bulk optical table \cite{armstrong2012programmable}, and Ref.~\cite{ferrini2013compact} analyzed the feasibility of such multipixel schemes. The present work moves that method onto a silicon-photonic chip with the same number of active pixels, performs the mixing as a physical RF network on the analog photocurrents so that a 32-channel array is read on two digitized channels, and expresses the feasibility condition of Ref.~\cite{ferrini2013compact} for a measured subset of a fixed detected subspace, with target homodyne angles and output calibration gains made explicit. Squeezing, displacement, and non-Gaussian resources remain optical. Optics generates and transports the quantum field \cite{vahlbruch2016detection, nehra2022few, BackerPeralCLEO2025}, electronics supplies stable, reprogrammable mode transformations \cite{GursesCLEO2024}, and the scheme applies to the Gaussian layer of measurement-based computation \cite{raussendorf2001one, van2007building, yukawa2008experimental, Jia2026}, to mode-selective detection \cite{fabre2020modes}, to CV quantum key distribution \cite{Weedbrook2012, grosshans2002continuous}, and to quantum sensing \cite{tse2019quantum}.

Optical coupling and detection loss limit the total efficiencies of the two halves to $4.1\%$ and $1.3\%$. Component-loss improvements demonstrated on this receiver platform give a path to sub-dB system loss \cite{Gurses2025}, at which the same RF network would recover the squeezing of the source rather than the few percent of it detected here, and the modular architecture allows the aperture to be scaled by duplicating channels.

With LO amplitude $\alpha_\text{LO}$, a quantum-limited coherent receiver produces the classical difference photocurrent $I(t) \propto |\alpha_\text{LO}|\,\hat{X}_\theta(t)$. The LO selects one optical quadrature and raises its fluctuations above the electronic noise, and photodetection destroys the optical field. Terminal protocols, including CV key distribution and cluster-state nullifier readout, require one selected quadrature per mode per shot, which a single-quadrature measurement interface provides at room temperature with $30.3$~dB of clearance.

A $193$-THz optical mode has $\hbar\omega/k_B \approx 9260$~K and thermal occupancy $\bar{n} \approx 4\times10^{-14}$ at room temperature, so thermal occupation of the optical mode is negligible, whereas a $5$-GHz microwave mode has $\hbar\omega_\mathrm{m}/k_B \approx 0.24$~K and $\bar{n} \approx 1.25\times10^{3}$. At room temperature the microwave vacuum lies $34$~dB below the thermal floor, so the electronics processes measured optical quadratures without resolving a microwave quantum state. Near $10$~mK the same mode has $\bar{n} \approx 4\times10^{-11}$ and superconducting parametric amplifiers approach the quantum limit \cite{macklin2015jtwpa}, so a cryogenic quantum-limited coherent receiver (below, cryogenic coherent receiver) with sufficiently low photodiode and first-stage noise would deliver quantum-noise-resolved RF quadrature statistics directly to superconducting qubits. Cryogenic photodiode responsivity and dark current, first-stage amplifier noise, and LO delivery without heat load set its engineering requirements.

Several architectures follow from a measurement interface of this kind. With a fixed $U_\text{opt}$, per-channel LO phases, and a reconfigurable real $U_\text{el}$, Eq.~(\ref{eq:condition}) defines a family of Gaussian circuits that is accessible with no optical reconfiguration. An $N$-mode programmable linear optical network needs $O(N^2)$ tunable optical elements and incurs their cumulative loss inside the quantum channel, whereas the equivalent RF network acts after detection and adds no optical loss, so the scaling variables become receiver channels and RF fan-in rather than the depth of the optical network \cite{GursesCLEO2024, Gurses2025, GursesCLEO2025, GursesQCT2026}. An adaptive measurement whose basis depends on an earlier outcome remains sequential and requires real-time LO control, modulation, or feedforward \cite{larsen2021deterministic}, since post-processing cannot change which noncommuting quadrature was measured.

A cryogenic coherent receiver co-located with superconducting circuits would extend this to circuits in which the electronic layer performs the feedforward. The nullifiers of an $N$-mode Gaussian cluster state are real linear combinations of quadratures and can be evaluated by an RF matrix multiplication on the photocurrent vector, reprogrammable from shot to shot with no optical element in the path, which Eq.~(\ref{eq:Mreal}) shows for $N=2$. Once the required quadratures have been measured, their linear combinations and the displacement feedforward can be implemented electronically, off-line when the protocol permits or at RF latency when it does not, and single-flux-quantum logic co-located with a cryogenic receiver could reduce that latency below the optical delay of the protocol. Gottesman--Kitaev--Preskill (GKP) stabilizers are displacement operators and their syndromes are modular quadrature variables \cite{gottesman2001encoding}, whose extraction requires the encoded ancilla interaction and measurement, after which electronics performs the linear combinations and modular decoding, so the receiver does not by itself implement GKP syndrome extraction.

Microwave--optical interfaces have transduced superconducting qubit states to optical photons \cite{mirhosseini2020superconducting}, read out a superconducting qubit over optical fiber \cite{Arnold2025}, entangled microwave and optical fields \cite{Sahu2023}, and transferred coherent signals between superconducting circuits in two dilution refrigerators over $1$~km of fiber \cite{Zhou2026}. In an electro-optic transducer, a pump detuned from the optical mode by the microwave frequency reduces the Pockels interaction to a beam-splitter coupling between one optical and one microwave mode \cite{tsang2010cavity}, with conversion efficiency set by the cooperativity and added noise set by pump-induced heating \cite{lauk2020perspectives, holzgrafe2020cavity}. A coherent receiver has the same frequency arrangement, with the LO at the pump frequency and the signal in a sideband at the microwave frequency, and differs in that photodetection folds the sideband beat into a classical photocurrent. With an electro-optic transducer in place of the balanced photodiodes of each channel and the LO as its pump, channel $j$ would deliver the microwave mode $\hat{c}_j = \sqrt{\eta_j}\,e^{-i\theta_j}\hat{b}_j + \sqrt{1-\eta_j}\,\hat{n}_j$, with $\eta_j$ the conversion efficiency, $\hat{n}_j$ the noise mode admitted by the converter, and $\theta_j$ the LO phase. The RF network would then act on bosonic microwave modes, so that a complex $U_\text{el}$ is a mode transformation, the reality condition, Eq.~(\ref{eq:condition}), is lifted while the support condition remains, and $W = U_\text{el}\,D(\vec{\theta})\,U_\text{opt}$ of Eq.~(\ref{eq:W}) becomes a general passive transformation of the converted modes, whose outputs are quantum states of the microwave field available to superconducting qubits. In this variant, conversion efficiency and added noise enter the quantum path once per channel, the loss of the superconducting microwave network attenuates the converted state, and the pump heat load per channel replaces the LO delivery requirement of the photodiode receiver. Teleportation-based transduction \cite{Zhong2020, Wu2021} and optically heralded entanglement \cite{Krastanov2021} provide alternative interfaces for quantum networks \cite{Kimble2008}. Optoelectronic quantum circuits of this kind, in which integrated photonics generates and distributes squeezing and entanglement, cryogenic electro-optic transducers transfer the optical modes into microwave modes, and superconducting microwave networks carry out the mode transformations on the converted states before they reach superconducting qubits, could couple quantum information between quantum optics and superconducting circuits.

These circuits could also be built in the opposite direction since the electro-optic interaction is bidirectional, and the same transducer with the same pump converts a microwave mode into the optical sideband \cite{tsang2010cavity, hease2020bidirectional, holzgrafe2020cavity}, $\hat{b}_j^{\prime} = \sqrt{\eta_j}\,e^{i\theta_j}\hat{c}_j + \sqrt{1-\eta_j}\,\hat{m}_j$, where the optical noise mode $\hat{m}_j$ admitted by the converter is in its ground state at any operating temperature. Run in this direction the receiver would be a coherent transmitter, an array of electro-optic modulators driven by the microwave modes with the LO as the carrier and the converted state in one sideband, with the optical network in transmission, $U_\text{opt}^\dagger$, returning the converted modes to the optical field. We call the microwave-to-optical direction electro-optic QIP. Optoelectronic QIP would transfer optical modes into the superconducting circuit and electro-optic QIP would transfer them back, so that a state enters the circuit for the deterministic non-Gaussian operations that the Josephson nonlinearity provides \cite{blais2021circuit}, which the Gaussian optical layer lacks \cite{menicucci2006universal}, and return to optics for distribution between cryostats, transport over fiber \cite{Zhou2026}, and room-temperature detection. Each transfer costs the conversion efficiency and added noise of the transducer once. Coupling both optoelectronic and electro-optic QIP could enable a photonic-electronic quantum computer, whose modes span the microwave and optical bands and in which squeezing and distribution are optical, nonlinearity and storage are superconducting, and mode transformations are carried out in either band.

\section*{Author contributions}

V.G. conceived the project and the idea, developed the theory, designed the measurement, built the experimental setup and carried out the measurements, wrote the analysis and figure code, analyzed the data, and wrote the manuscript. A.H. supervised the work. Both authors discussed the results and revised the manuscript.

\section*{Data availability}

The reduced products underlying Fig.~\ref{fig:data} (phase-binned moments, fitted coefficients, reconstructed covariance matrices, bootstrap samples, the surrogate null, and the band-assignment scan) are provided as an ancillary file with the arXiv version of this article (\texttt{anc/fig3data.npz}). The raw $20$-MSa/s two-channel acquisitions are $370$~MB each and are available from the corresponding author on reasonable request.

\section*{Code availability}

The figure-generation code is distributed with this article in \texttt{anc/}. It regenerates Fig.~\ref{fig:data} from the reduced products and prints the quantities quoted in Sec.~\ref{sec:results} and Appendix~\ref{app:stats}.

\appendix

\section{Derivation of the measurement relation}
\label{app:derivation}

Substituting $\hat{b}_j = \sum_n (U_\text{opt})_{jn}\hat{a}_n$ into Eq.~(\ref{eq:quadrature}) and then into Eq.~(\ref{eq:electronic}) gives
\begin{equation}
    \hat{Y}_k = \frac{1}{\sqrt{2}} \sum_n \left[ W_{kn}\, \hat{a}_n + \overline{\widetilde{W}_{kn}}\, \hat{a}_n^\dagger \right],
    \quad
    \widetilde{W} = \overline{U_\text{el}}\, D(\vec{\theta})\, U_\text{opt},
    \label{eq:expand}
\end{equation}
where the overbar denotes complex conjugation. The coefficients are conjugates, $\widetilde{W} = W$, when $U_\text{el}$ is real, and $\hat{Y}_k$ then takes the single-quadrature form of Eq.~(\ref{eq:measurement}) with $\hat{c}_k = \sum_n W_{kn}\hat{a}_n/\norm{\vec{w}_k}$. The commutator
\begin{equation}
    [\hat{c}_k, \hat{c}_l^\dagger] = \frac{(WW^\dagger)_{kl}}{\norm{\vec{w}_k}\,\norm{\vec{w}_l}}
\end{equation}
equals unity for $k = l$ by the choice of the Euclidean norm and vanishes for $k \neq l$ exactly when the rows of $W$ are orthogonal. In general $[\hat{X}_{c_k}, \hat{X}_{c_l}] = i\,\text{Im}\,(WW^\dagger)_{kl}/(\norm{\vec{w}_k}\norm{\vec{w}_l})$. Here $WW^\dagger = U_\text{el}D(\vec{\theta})U_\text{opt}U_\text{opt}^\dagger D(\vec{\theta})^\dagger U_\text{el}^T = U_\text{el}U_\text{el}^T$ is real because $U_\text{el}$ is real and the rows of $U_\text{opt}$ are orthonormal, so $[\hat{X}_{c_k}, \hat{X}_{c_l}] = 0$ and the $\hat{Y}_k$ admit a joint distribution whether or not the virtual modes are orthogonal.

\section{Reconstruction and statistical analysis of Fig.~\ref{fig:data}}
\label{app:stats}

\subsection{Acquisition and calibration}

Both combined photocurrents are digitized simultaneously with a $100$-MSa/s digitizer and stored at $20$~MSa/s for $6.5$~s with a $0.3$~V full scale. The acquisition is divided into $495$ blocks of $13.1$~ms, within each of which the two channels are demodulated into nine $1$-MHz bands spanning $0.5$--$9.5$~MHz and the $4\times4$ second-moment matrix of $(\hat{X}_{s_1}, \hat{P}_{d_1}, \hat{X}_{s_2}, \hat{P}_{d_2})$ is accumulated. Bands $0.5$--$2.5$ and $6.5$--$9.5$~MHz form the phase reference, and the four bands centered at $3$--$6$~MHz ($2.5$--$6.5$~MHz) the tested covariance.

Each quadrature is normalized to the mean vacuum marginal of the two quadratures of its own array half, taken with the squeezed beam blocked, and each cross-half moment to the geometric mean of the two. The two acquisitions were logged at LO photocurrents of $-4.24$ and $-4.27$~dB relative to the reference monitor, so the vacuum level, which scales linearly with LO power, is scaled by the corresponding $0.03$~dB before normalization. The photocurrent log has $0.01$~dB resolution. The correction is an experimental calibration between two acquisitions taken at different LO powers and not a fitted parameter. We use one full log increment, $\pm0.01$~dB, as a systematic allowance on the between-acquisition LO correction. A perturbation $\delta$ in dB to the applied correction rescales the covariance as $\Sigma(\delta)=10^{-\delta/10}\Sigma(0)$ and rescales every ordinary and partial-transpose symplectic eigenvalue by the same factor. To first order, the corresponding allowance is $u_\mathrm{cal}=(\ln 10/10)\tilde\nu_-\times0.01$, giving $u_\mathrm{cal}=0.00229$ for the symmetric sideband modes. It is quoted separately from the bootstrap standard deviation $s_\mathrm{stat}=0.00081$. The nominal significance uses $(1-\tilde\nu_-)/\sqrt{s_\mathrm{stat}^2+u_\mathrm{cal}^2}$. A vacuum-reference error rescales $\nu_-$ and $\tilde\nu_-$ together, by $7\times10^{-4}$ for $0.003$~dB, and an error of $0.0034$~dB brings $\nu_-$ to unity while leaving $\tilde\nu_- = 0.9919$.

The recovered phase $\psi$ is the argument of the analytic signal of the common-mode reference-band power, bandpassed to $0.8$--$6$~Hz around twice the ramp frequency. The reference bands are disjoint from the bands that enter $\Sigma$, and the LO phase is common to every sideband frequency while the quantum noise is not, so the reference is statistically independent of the tested covariance.

\subsection{Covariance reconstruction}

Each of the three independent second moments is regressed on $(1, \cos\psi, \sin\psi)$ over all $495$ blocks by ordinary least squares, giving the coefficients $a$, $b$, and $c$ of Eq.~(\ref{eq:tomo}). For the symmetric sideband modes, the fit to half 1 gives $\Sigma_{11}=a_1+b_1$, $\Sigma_{22}=a_1-b_1$, and $\Sigma_{12}=c_1$, with the analogous relations for half 2, and the cross-half fit gives $\Sigma_{13}=a_{12}+b_{12}$, $\Sigma_{24}=a_{12}-b_{12}$, and $(\Sigma_{14}+\Sigma_{23})/2=c_{12}$, where the constant $a_{12}$ is the electronic offset described below and is set to zero. Equation~(\ref{eq:tomo}) therefore constrains nine independent linear combinations of the ten covariance parameters. The covariance matrix is assembled from these entries and symmetrized. Symplectic eigenvalues are the moduli of the eigenvalues of $i\Omega\Sigma$ with $\Omega$ the standard symplectic form, and the partial transpose flips the sign of $\hat{P}_2$.

The component of each cross-half moment that does not rotate with the recovered phase is an electronic offset and is set to zero before the optical covariance is assembled. The offset is $-0.14689$ for the reported $3$--$6$~MHz test bands of the squeezed acquisition and $-0.02600$ for the vacuum acquisition. Retaining the offset of the squeezed acquisition gives $\nu_-=0.85431$, so that uncorrected second-moment matrix is not a physical optical covariance. The range of subtraction fractions that yields a physical covariance matrix begins at $0.99242$, and from that boundary to the applied correction $\tilde\nu_-$ changes from $0.99269$ to $0.99270$.

The ramp constrains $\langle\hat{X}_{s_1}\hat{P}_{s_2}\rangle+\langle\hat{P}_{s_1}\hat{X}_{s_2}\rangle$ but not the antisymmetric combination, which is set to zero. Adding $\epsilon$ to $\Sigma_{14}$ and subtracting it from $\Sigma_{23}$ over $\epsilon = \pm0.01$, five times the largest measured $\hat{X}\hat{P}$ entry $\Sigma_{34} = 0.0020$, changes $\tilde\nu_-$ by less than $10^{-4}$, whereas $\nu_-$ falls below unity beyond $\pm0.0012$.

\subsection{Uncertainties and validation}

Physicality is assessed at the nominal calibration. Of the $600$ original bootstrap estimates, $581$ for the symmetric and $594$ for the antisymmetric sideband modes satisfy $\nu_-\geq1$. All resamples are retained in the quoted intervals. The smallest symplectic eigenvalue $\nu_-$ of the untransposed covariance matrix for the symmetric sideband modes has a $95\%$ bootstrap percentile interval $[0.99996,1.00202]$, so physicality is not resolved throughout the uncertainty interval.

Statistical uncertainty is estimated with a moving-block bootstrap using $600$ resamples. Each resample concatenates randomly selected overlapping sequences of $25$ consecutive acquisition blocks and is truncated to the original length of $495$ blocks. The second moments and their associated recovered phase values are resampled together, and the covariance fit is repeated. The vacuum normalization, the LO correction, the electronic-offset model, and the recovered phase reference are held fixed across resamples, so the interval is statistical only. For the surrogate validation test, a common random Fourier phase is added at each frequency to all three second-moment time series. This preserves their power spectra and cross-spectra while disrupting their phase alignment with the reference. The estimator is evaluated for $400$ such surrogates against the unchanged reference phase.

The reference/test split is not unique, so every assignment with a contiguous test block of two to four bands and a reference of three or four consecutive bands among the remaining bands is reconstructed and tested, $151$ in all. The reported assignment is the center of the analysis band, chosen for flatness away from both filter edges, and uses a five-band reference outside this family. Its nearest scan value, $0.99270$, is the $20$th of $151$ in rank order, at the $13$th percentile.

Unequal overlap efficiencies of the two halves give the ideal ratio of cross-half to single-half modulation amplitude $\sqrt{\eta_1\eta_2}/\bar{\eta}\leq1$. The fitted ratio for the reported bands is $1.11$, with a $95\%$ interval $[0.98,\,1.27]$ from propagating the binned standard errors, consistent with the model bound.

\subsection{Gaussian loss model}

In a Gaussian loss model in which every imperfection is equivalent to an additional loss $1-\eta$ at the source of the squeezed light \cite{Asavanant2019}, the fitted mean excess $\bar{V}-1$ and peak-to-peak modulation depth $D$ of the single-half variance satisfy $(\bar{V}-1)/D = \tanh(r)/2$ for squeezing parameter $r$, and the two halves invert to $r = 0.10$ and $0.26$ with $\eta = 4.1\%$ and $1.3\%$. The spread between the halves follows from a $10^{-3}$ difference in fitted mean excess, so the inversion is ill-conditioned.

\bibliography{references}

\end{document}